\documentclass[conference]{IEEEtran}
\IEEEoverridecommandlockouts
\usepackage{amsmath,amssymb,amsfonts}
\usepackage{algorithmic}
\usepackage{algorithm}
\usepackage{graphicx}
\usepackage[numbers,sort&compress]{natbib}
\usepackage{url}
\usepackage{textcomp}
\usepackage{xcolor}
\def\BibTeX{{\rm B\kern-.05em{\sc i\kern-.025em b}\kern-.08em
    T\kern-.1667em\lower.7ex\hbox{E}\kern-.125emX}}
    
\begin{document}
\title{Spotting Fake Profiles in Social Networks via Keystroke Dynamics}

\author{ 
\IEEEauthorblockN{Alvin Kuruvilla}
\IEEEauthorblockA{\textit{Department of Computer Science} \\
\textit{Hofstra University, NY, USA}\\
akuruvilla1@pride.hofstra.edu}
\and
\IEEEauthorblockN{Rojanaye Daley}
\IEEEauthorblockA{\textit{Department of Computer Science} \\
\textit{Hofstra University, NY, USA}\\
daley.rojanaye@gmail.com}
\and
\IEEEauthorblockN{Rajesh Kumar}
\IEEEauthorblockA{\textit{Department of Computer Science} \\
\textit{Bucknell University, PA, USA}\\
rajesh.kumar@bucknell.edu}
}

\maketitle

\begin{abstract}
Spotting and removing fake profiles could curb the menace of fake news in society. This paper, thus, investigates fake profile detection in social networks via users' typing patterns. We created a novel dataset of $468$ posts from $26$ users on three social networks: Facebook, Instagram, and $X$ (previously Twitter) over six sessions. Then, we extract a series of features from keystroke timings and use them to predict whether two posts originated from the same users using three prominent statistical methods and their score-level fusion. The models' performance is evaluated under same, cross, and combined-cross-platform scenarios. We report the performance using k-rank accuracy for $k$ varying from $1$ to $5$. The best-performing model obtained accuracies between $91.6\%-100\%$ on Facebook (Fusion), $70.8-87.5\%$ on Instagram (Fusion), and $75\%-87.5\%$ on X (Fusion) for $k$ from $1$ to $5$. Under a cross-platform scenario, the fusion model achieved mean accuracies of $79.1\%-91.6\%$, $87.5\%-91.6\%$, and $83.3\%-87.5\%$ when trained on Facebook, Instagram, and Twitter posts, respectively. In combined cross-platform, which involved mixing two platforms' data for model training while testing happened on the third platform's data, the best model achieved accuracy ranges of $75\%-95.8\%$ across different scenarios. The results highlight the potential of the presented method in uncovering fake profiles across social network platforms. \footnote{©2023 IEEE. Personal use of this material is permitted. Permission from IEEE must be obtained for all other uses in any current or future media, including reprinting/republishing this material for advertising or promotional purposes, creating new collective works, for resale or redistribution to servers or lists, or reuse of any copyrighted component of this work in other works. To appear in 2024 IEEE 21st Consumer Communications \& Networking Conference (CCNC)}
\end{abstract}


\section{Introduction}
\label{Introduction}
\subsection{Social networking platforms}
Social networking platforms such as Facebook, Instagram, Twitter, TikTok, Pinterest, and TruthSocial have become indispensable. They affect almost every aspect of our lives, from politics, economy, world peace, public health, and mental health. These platforms enable us to post content and share information virtually with a set of people or publicly. Facebook, Instagram, and Twitter arguably have the largest number of active users \cite{Gongane2022}. Facebook allows users to create posts of $63000$ characters, share, and react to them. Some platforms, for instance, Facebook, have evolved and offer users opportunities to create and react on social forums. In addition, it encourages people to trade new or used products with relative ease. Instagram is more focused on sharing pictures and videos, allowing users to write captions of up to $2000$ characters and comment on posts. Twitter is dedicated to concise text posts. These short posts, called tweets, can have up to $280$ characters (previously $140$ characters) and one to four pictures \cite{TwitterElon280}. Most users use Twitter to keep up with the news \cite{TwitterGen}. Recently, Twitter has become influential in politics and the financial investment community \cite{TwitterFinance}.

The increasing integration of social media into daily life has amplified the risks posed by fake profiles \cite{LiteratureReview2023}. The Pew Center reports that $84\%$ of adults aged $18-29$ are active on social networks, with significant usage also seen in older age brackets, with $81\%$ for adults ages $30-49$, $73\%$ for those aged $50-64$, and $45\%$ for those $65$ and older \cite{Pew2021}. Despite Facebook's dominance as a leading platform used by $69\%$ of the surveyed American adults, as of a 2021 Pew Center study, its growth has stagnated since 2019. Meanwhile, platforms like Instagram and Twitter command attention, with $40\%$ and $23\%$ regular users, respectively. The widespread adoption of these platforms underscores the pressing concern about fake profiles and their potential influence.

\subsection{Fake profiles, their use, and misuse}
Across a multitude of social network websites, millions of posts are written and shared every day \cite{LiteratureReview2023}. Users are encouraged to make accounts to share new stories, videos, and opinions. An online profile is an extension of the self. Ideally, users would treat each other with the same kindness and decorum online as they would if interacting in person. Some users may create additional profiles on the platform to provide tailored content to their followers. Users may make an additional Instagram account dedicated to photos of their dog or a secondary Twitter account to discuss their favorite band. Individuals with multiple accounts encourage their followers to follow them on both accounts, as it is easy to log into multiple accounts and switch between them. Users can link their accounts across multiple platforms using the same email. There is an online, usually public, record of every user interaction. The public nature encourages users to make additional accounts with anonymous or false identities to engage in publicly immoral behavior and share controversial and hateful opinions. Creating an account with a separate identity protects users' reputations, enabling them to post freely without personal repercussions. 

Fake profiles damage users across all social network platforms \cite{LiteratureReview2023}. Mauro et al. \cite{Mauro2012} describe two major fake profile attacks. The first is an Identity Cloning Attack (ICA), in which a user's information is stolen and used to make any number of clone accounts, each pretending to be the victim's original account \cite{Mauro2012}. The second attack is the Fake Profile Attack (FPA), where a user's personal information is taken and uploaded to a site where the user does not have an account \cite{Mauro2012}. Both attack styles allow the clones to befriend the victim's network to launch more attacks, compromising the victim's credibility \cite{Mauro2012}. These fake accounts harm the users' reputations and may steal data and flood other users with spam or phish them for information. Fake profiles pose a threat to both users and the social network platforms. Khaled et al. \cite{Khaled2018} noted that these fraudulent accounts erode trust in user metrics, adversely affecting advertisers' and developers' revenue streams. In a related trend, U.S. banks have recently started examining loan applicants' Twitter and Facebook profiles before approving their applications \cite{Khaled2018}. Thus, the repercussions of fake profiles on social networks extend to institutions across various sectors relying on social network data for assessments, potentially distorting their judgments based on inaccurate information influenced by these fake profiles.

When social network users share fake news stories and hateful content online, it is vital to hold those individuals accountable. For example, a former professor of chemistry at the University of New Hampshire used multiple Twitter accounts to share hateful rhetoric online \cite{UNH}. On his personal Twitter account, the then-professor was courteous and professional. However, on the secondary, anonymous Twitter account, the Caucasian professor posed as a woman of color and spoke out against racial justice and gender nonconformity and harassed other users online. The account was created in January $2019$ and remained active until October $2020$, amassing more than $13,000$ followers before the identity of the account was revealed. Following public outcry, the professor removed himself from the university \cite{UNH}. If systems existed to compare these two accounts' typing or speech patterns, the link between them could have been detected earlier, significantly undermining the false account's credibility and influence.

\subsection{Keystroke dynamics}
Keystroke dynamics, also known as keystroke biometrics, use the timings of key press and release on a keyboard \cite{shadman2023keystroke}. Typing speeds and patterns vary enough between individuals to make biometric user identification feasible. Fixed text and free text are the two main types of keystroke dynamics . Ayotte et al. \cite {FastFreeText} state that fixed text requires the test sample keystrokes to match the profile keystrokes. On the other hand, the free text puts no restrictions on the keystrokes users can type.

Keystroke dynamics provide a unique opportunity for user identification, authentication, and authorship of posts \cite{ali2017keystroke}. Social network users can share large amounts of text on platforms with their connected network of followers. These typing and speech patterns can identify individuals based on their typing behavior, like fingerprints or other biometric identifiers. 

\subsection{Social networks and keystroke dynamics}
The benefit of connecting social network platforms and keystroke dynamics lies in their potential collaboration to enhance security and address user identification issues, such as unauthorized account access. Suppose the typing patterns of an account suddenly change, indicating a potential compromise. The platform can prompt users for additional authentication or take other security measures to prevent unauthorized access. Beyond user authentication and account safeguarding, keystroke biometrics serve as a proactive fraud detection tool. Scammers and malicious actors frequently target social network platforms. Keystroke dynamics can help identify anomalous behavior that might indicate fraudulent activities, such as a compromised account. Like any biometric system, however, false positives are possible, denying legitimate users access, or false negatives, where unauthorized users can gain access. While keystroke dynamics could offer significant benefits, managing the potential trade-off between precision and user experience is critical to ensure effectiveness.
\subsection{Main contributions} Our contributions are manifold, as listed below: 

\begin{itemize}
\item Developed a Python-based key logger that operates in the background. This key logger captures all keystrokes a user makes while it's active. The code has been made publicly available to support and encourage research. 
\item Established a new dataset containing keystrokes from Facebook, Twitter, and Instagram. This data was gathered over six sessions and involved 26 users. Accompanying this paper, the dataset will be publicly available, inviting the research community to utilize it and potentially expand its user base.
\item Tested and compared the efficiency of three well-known keystroke verification methods, one considered state-of-the-art. The comparison spanned same, cross, and combined-cross platform scenarios.
\item Conducted an analysis on the effectiveness of score-level fusion using the three different classifiers to improve overall detection performance.
\item We make the dataset and code available for future research in this direction. \footnote{https://github.com/AlvinKuruvilla/social-network-fake-profile-detection.git}. 
\end{itemize}

The structure of this paper is outlined as follows: In Section \ref{sec:RelatedWorks}, we review relevant literature and previous studies. Section \ref{ProposedModel} introduces our proposed framework. Our methods for data collection and feature extraction are detailed in Sections \ref{DataCollection} and \ref{FeatureExtraction}, respectively. The evaluation scenarios for our model are elaborated upon in Section \ref{ModelEval}, while Section \ref{PerformanceEval} explains the metrics employed to assess the model's efficacy within the three scenarios. Section \ref{ResultsAndDiscussions} is dedicated to discussing our findings. We summarize our work and suggest avenues for future research in Section \ref{ConclusionAndFutureWork}.

\section{Related work}
\label{sec:RelatedWorks}
\subsection{Keystroke applications}
Keystrokes have been widely studied for user authentication \cite{BelmanTOPS, Kumar2016BTAS, shadman2023keystroke}. We propose keystroke biometrics as a basis for fake profile detection as keystroke is an established method of user identification and authentication across devices and platforms \cite{Kumar2016BTAS, BelmanTOPS}. Also, keystroke has been used for authorship attribution \cite{plank-2018-predicting}. Monaro et al. \cite{Monaro2018CovertLD} used free text keystroke data to determine whether or not participants were lying in their written responses. Their study utilized $40$ participants for model training and 10-fold cross-validation. The model was tested online with an additional $151$ participants, $86$ of whom were in the liar category. Monaro found that liars exhibited longer reaction times at the onset of responding than those answering questions authentically \cite{Monaro2018CovertLD}. Keystroke dynamics have also revealed users' soft biometrics such as gender and handedness \cite{Agrawal2020}.  

\subsection{Fake profile and content moderation in social networks}
There exists a mechanism to moderate content people post on social networks. Each social network site has post regulations or community guidelines to deter harmful content online and a tiered punishment system. On social networks, accounts with fewer followers or posts that receive fewer likes or shares may share harmful content without consequences until another user reports the post or the account \cite{ContentModDefamation}. After a first offense, the post is deleted. Repeated violations end in account suspension and deletion. Users may find that their account has been shadow-banned: posts or even their account gaining less attention after being suppressed by moderators. Although that profile has been punished hypothetically, creating a new profile with a new email account and starting a new one is easy. Many social network sites, such as Facebook and Instagram, try to deter bots and harassment accounts by quickly monitoring the number of posts made. An account that posts too often or tags the same users too often could be fake. Some sites may even limit the number of posts made in 30-minute intervals. Sites also hire moderators to review accounts and posts that have been reported, but it may take a few days before the posts are even reviewed \cite{Gongane2022}. 

\subsection{Fake profile and keystrokes in social networks}
The most closely related work is by Morales et al. \cite{Morales2020} and Bhattasali et al. \cite{Bhattasali2021}, who have used keystrokes for fake account detection. However, they use datasets that do not mirror the real-world social network context. For example, the dataset used in \cite{Morales2020} used the Aalto University dataset, which required subjects to memorize English sentences and then type them as quickly and accurately as possible. This is different from how people type on social networks. On the other hand, Bhttasali et al. \cite{Bhattasali2021} proposed $DEEP\_ID$, which utilizes keystrokes, mouse clicks, and touch strokes of the users to flag fake user accounts silently. However, to our knowledge, the dataset and code are not available for future investigations. Realizing there exists no publicly available dataset, to the best of our knowledge, we created a new dataset that closely mirrors how users make posts or comments on social networks. We also used traditional classifiers and their fusion, which can be scaled to many users. 

\subsection{Other fake profile detection methods}
In addition to monitoring an account's age and post frequency, another way to verify an account is by analyzing the network of connections between profiles. Fake profiles often have limited or shallow connections with other users\cite{Mauro2012}.
According to Gongane et al. \cite {Gongane2022}, Natural Language Processing (NLP) with Machine Learning (ML) algorithms and Deep Neural Networks are deployed to detect and moderate detrimental content on social network platforms. Some machine learning tools and bots can remove banned content; however, they lack the complexity to understand the grammatical, cultural, or journalistic context. In addition, these bots require testing and training data to inform their decisions. Although there may be plenty of data, the data is built on previous policies and does not account for rapid policy change \cite {ContentModAI}.

Khaled et al. \cite{Khaled2018} utilized categorical features, e.g., language, profile-sidebar-color, tweets,  friends-count, followers-count, default-profile, image profile-use-background, etc.

\subsection{The gap}
While previous research has extensively examined keystrokes for purposes such as authorship attribution, authentication, and identification, this study uniquely focuses on utilizing typing patterns to detect fake profiles. A notable void in previous literature is the need for datasets closely mirroring the behavior of users on social networks. Our hypothesis derives from the understanding that users' typing patterns may vary due to context (such as walking, traveling, sitting, or sleeping) or the specific platform in use (text editors, code editors, social media platforms).

Furthermore, the complexity of keystroke patterns can be influenced by multiple factors, including the content being typed, the language used \cite{multi_keyboard_lanaguage}, the mode of typing (free or fixed-text \cite{FastFreeText, SimCVPRDigraphs}), the user's emotional state \cite{emotional_state_keystroke}. Even the device, keyboard \cite{multi_keyboard_lanaguage}, or application employed for typing could impact the keystrokes. Certain contextual elements exert a more pronounced impact on keystroke patterns than others. For instance, while a keyboard change might have a relatively minor effect on keystroke patterns, shifts in mental or emotional states can lead to more substantial variations \cite{emotional_state_keystroke}.

Given the absence of suitable data tailored to our problem statement, we curated a dataset within an unrestrained environment and real social network platforms such as Facebook, Twitter, and Instagram. Our approach was designed to scale to billions of users. Consequently, we employed established keystroke verifiers and their fusion. Notably, the fusion of these verifiers consistently outperformed individual algorithms across most experimental scenarios.

\section{Experimental details}
\subsection{Proposed framework}
\label{ProposedModel}
\begin{figure}[htp]
    \centering
    \includegraphics[width=3.4 in, height = 1.45in]{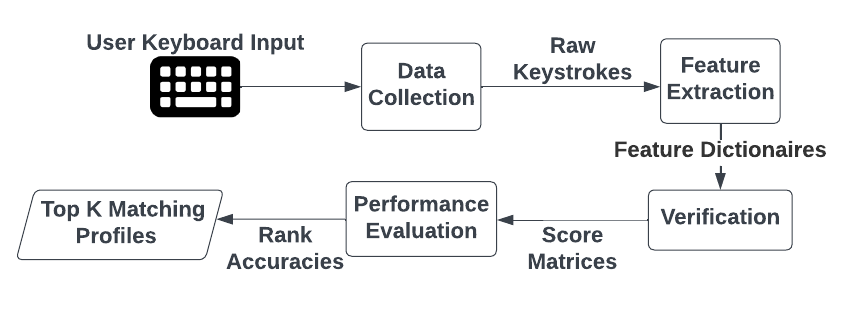}
    \caption{Our proposed framework, as depicted, involves collecting participants' keystrokes and deriving feature dictionaries. The feature dictionaries are fed to the keystroke verifiers, resulting in score matrices. The performance evaluation protocol selects the top $k$ profiles from these matrices.}
    \label{fig:MLPipeline}
 \end{figure}

The proposed framework is presented in Figure \ref{fig:MLPipeline}. The major steps are explained below:
\begin{enumerate}
    \item Collect keystroke data while users make posts on social network platforms.
    \item Extract unigraphs, digraphs, and word-level features \cite{Agrawal2020, BelmanTOPS, Kumar2016BTAS, SimCVPRDigraphs}. 
    \item Feed the features to the three verifiers to get the corresponding match scores, and then take the mean and median of the score matrices to compute the fusion score matrices \cite{FastFreeText, SimCVPRDigraphs, BelmanTOPS}.  
    \item Compute k-rank accuracy from the score matrices.
\end{enumerate}

These steps are described in detail in the following sections. 

\subsection{Data collection}
\label{DataCollection}
\subsubsection{Why a new dataset?}
Social network posts are typically a person's response to some stimuli, media, or event. Thus, to emulate the posting experience on social networks, the participants were asked to watch three videos that would stimulate their minds and initiate a response. We chose these videos to facilitate a possible emotional response from participants and make them interested enough to type their thoughts down, similar to how people generally post on social media.

Several keystroke datasets are freely available, designed specifically for authentication and identification purposes with free and fixed text. However, none of the existing datasets consisted of keystrokes captured on a social network platform that reflects real users' engagement with typical activities performed on such networks, including watching videos, commenting, or creating entirely new posts. We formulated our data collection protocol to address this and received approval from the University's Institutional Review Board (IRB). Though numerous social network platforms are available, our experiment concentrated on the most widely-used public platforms. Facebook, Instagram, and Twitter are the most heavily used and common ones that have fake profile issues \cite{PopularSiteProblemOfFakeAccounts, Gongane2022}.

\subsubsection{Data collection pipeline}
\begin{figure}[htp]
    \centering
    \includegraphics [width=3.3 in, height = 1.2in] {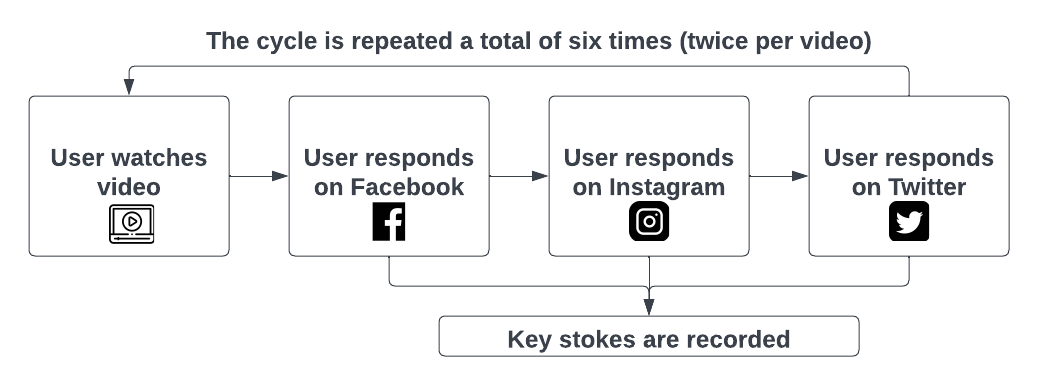}
    \caption{The data collection pipeline. The user watches one of the videos and posts on each social network platform (using a dummy account we set up). The entire pipeline is run six times (twice per video)}
    \label{fig:DataCollectionSetup}
\end{figure}

Figure \ref{fig:DataCollectionSetup} provides an overview of the data collection pipeline. As demonstrated, users watched videos and posted their thoughts on Facebook, Instagram, and Twitter (using a dummy account we set up). They could watch them as often as needed, which we believe is a natural behavior. The procedure was repeated twice per video. Metadata about the participants, such as age, gender, handedness, and education level, was also collected. 

The following are the video clips the participants watched during the data collection experiment. 
\begin{enumerate}
   \item A clip from the 2005 movie \textit{Coach Carter} \cite{DataCollectionVideo1}.
  \item A clip of Chris Rock and Will Smith's altercation during the 2022 Oscars \cite{DataCollectionVideo2}. 
  \item A video excerpt of President Joe Biden's speech to NATO regarding the Russia-Ukraine war from March 26, 2022 \cite{DataCollectionVideo3}.
\end{enumerate}

Participants were invited via emails, flyers, and word of mouth. They booked appropriate slots via Calendly. $32$ people signed up, and $26$ users' data was collected that we used in our experiment. Each user watched the three videos twice, resulting in six different sessions. Each session was separated by at least five minutes. Most of the participants completed the exercise on the same day. However, some participants came the next day for the last three sessions.

The participants were briefed about the experiment, its hypothesis, and the activities involved in the data collection. They then signed the consent form approving their data collection and use of data for research purposes. 

\subsubsection{Data collection app}
The keystroke data was collected using an in-house Python-based key logger controllable from the laptop terminal ($2020$ Apple MacBook Pro with $2.3$ GHz Intel Core i7, $8$GB RAM) \cite{KuruvillaKMLog}. Although we used a MacBook Pro, the key logger can collect data from any computer. We have open-sourced the software for other people to use.

\subsubsection{Data collection environment}
\begin{figure}[htp]
     \centering
     \includegraphics[width=3.3 in, height = 2 in]{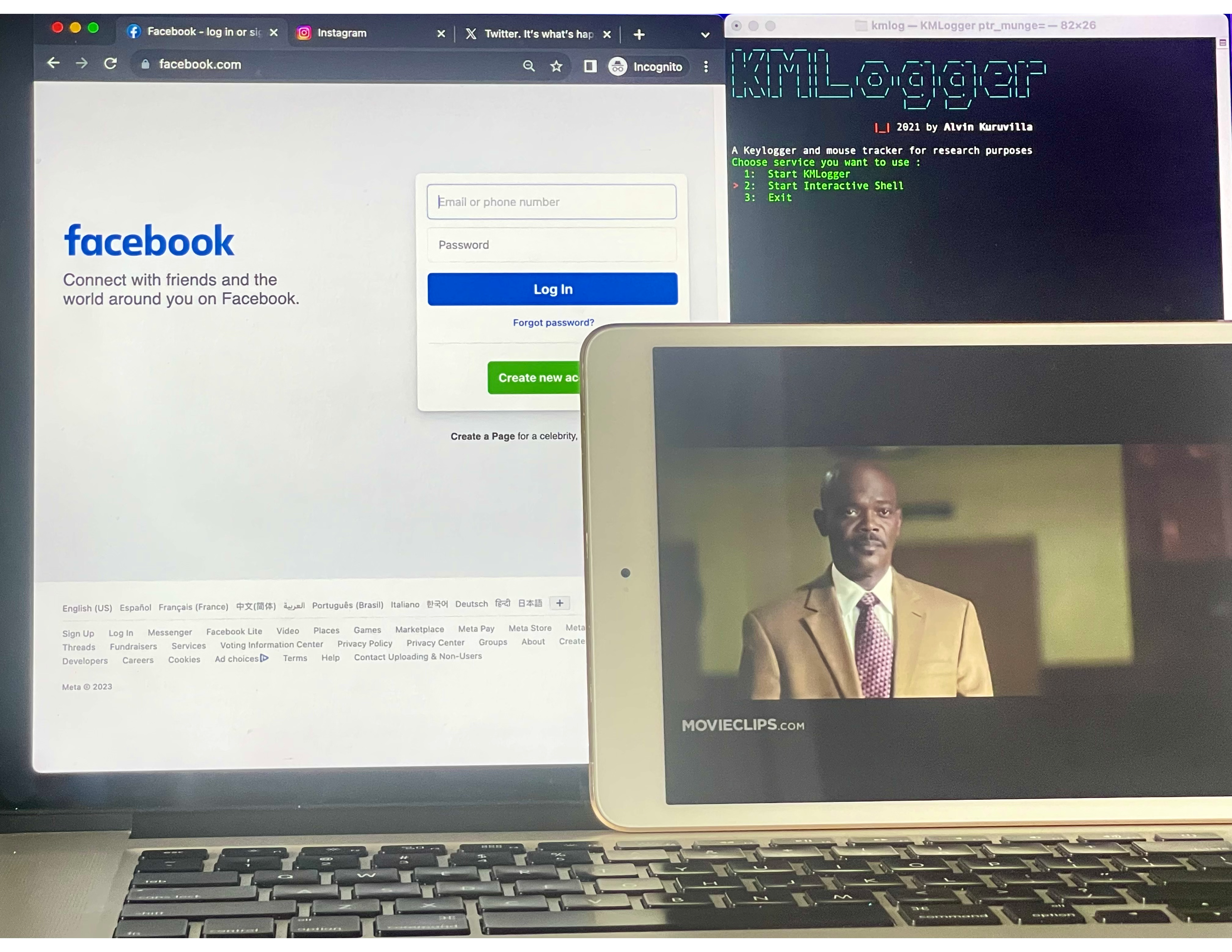}
     \caption{An illustration of our data collection environment. The user would watch a video and post their thoughts on social media while the key logger (top right) continuously collected their keystrokes in the background.}
     \label{fig:DataCollectionExample}
 \end{figure}

Figure \ref{fig:DataCollectionExample} demonstrates the data collection environment. As seen in Figure \ref{fig:DataCollectionExample}, we used the actual Facebook (and Instagram and Twitter) interface to collect participant data. 

\begin{figure*}[htp]
    \centering
    \includegraphics[width=7.1 in, height=1.5in]{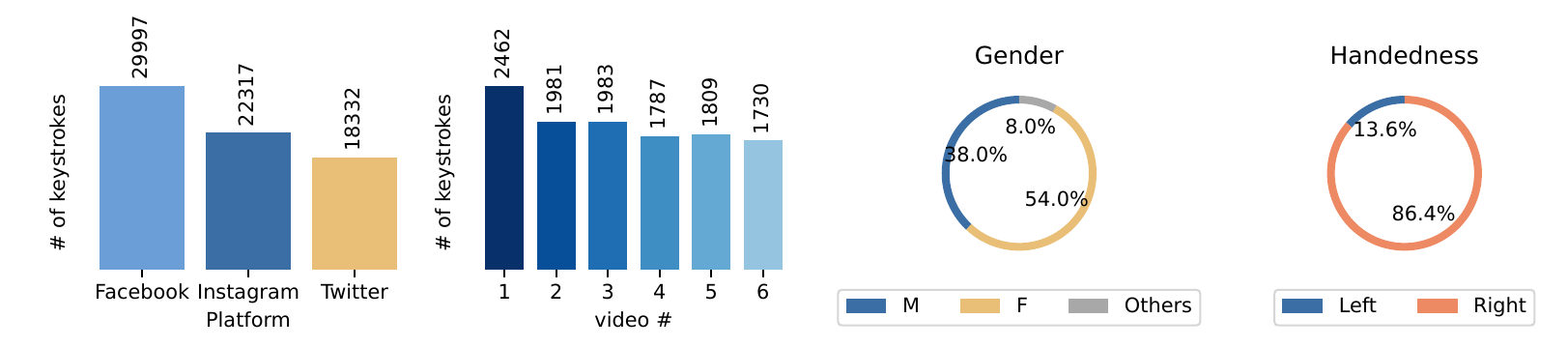}
    \caption{Stats of keystrokes generated by all participants for different platforms, videos, and the summary of participants' demographics. }
    \label{fig:datsetstats}
\end{figure*}

\subsubsection{Data collection statistics} 
\begin{figure}[htp]
    \centering
    \includegraphics[width=3.5 in, height=2.4in]{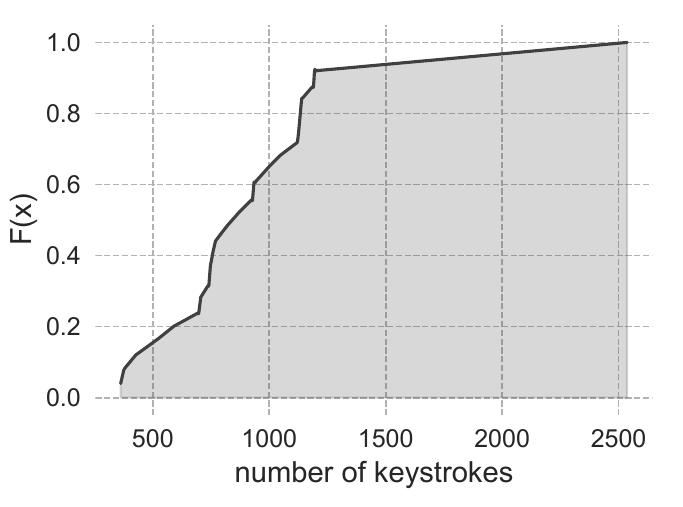}
    \caption{Empirical cumulative distribution function (ecdf) demonstrating the number of keystrokes for users in the dataset.}
    \label{fig:keystroke_ecdf}
\end{figure}

Figures \ref{fig:datsetstats} and \ref{fig:keystroke_ecdf} illustrate dataset statistics. Facebook comments were much larger in length, followed by Instagram and Twitter. Unsurprisingly, amount of keystrokes generated declined as the user moved toward the sixth session. The dataset consisted of 54\% female and 38\% male participants, while 8\% were non-binary. 86\% of people were right-handed, while the rest were left-handed. Figure \ref{fig:keystroke_ecdf} shows the empirical cumulative distribution of the number of keystrokes across the users. We can see that some participants expressed their thoughts in fewer words than others. 

\subsection{Feature extraction}
\label{FeatureExtraction}
We extracted unigraphs (Key Hold Time), digraphs (Key Interval Time), and word-level features from our dataset. These features are described below:
\begin{itemize}
  \item \textit{Unigraphs} is calculated by finding the difference between key press and release times for each occurrence of a key. For example, if the letter 'b' was pressed and released $20$ times, the unigraphs for this 'b' would be a list of $20$ elements such that each element is the difference between the corresponding key release and key press events.
  \item \textit{Digraphs} capture press and release timings for two consecutive keys, $k_{i}$ and $k_{i+1}$. There are four possibilities, as we have press and release events associated with each key. Assuming strong correlations among the four possibilities would exist, we used just one of the four possibilities described below. Another motivation was to keep the model simple and scalable to large social networking platforms. 
  \begin{equation}   
     F_1 = k_{i+1} (\text{press time}) - k_{i} (\text{release time})
  \end{equation} 
  \item \textit{Word-graphs} Several previous studies have recommended word-level features in the context of free-text typing \cite{BelmanTOPS, SimCVPRDigraphs, Agrawal2020}. Specifically, we used the word hold time as described in \cite{BelmanTOPS, Agrawal2020}. 
\end{itemize}

The extracted features were stored in a dictionary format where the dictionary keys represented the key, pair of keys, or word typed, and the corresponding values were a list of the feature values for each occurrence of the key, pair of keys, or word. This was done for each user, each platform, and each session.

\subsection{Model Evaluation Setup}
\label{ModelEval}
\subsubsection{Training and testing data}
\label{TrainingAndTestingDataSplit}
We separated the data for training and testing using session IDs. A session corresponds to the duration of watching one video. By this metric, we had six sessions where the first three videos were the same as the last three videos. However, session-level video ordering was not enforced as we wanted to keep our method independent of the videos watched and the content typed. Since we had three social networking platforms, we tested the proposed framework under the following scenarios. 

\textit{Same platform} For each user, the data collection for a particular platform was divided into two parts. The first part contained data collected during the first three sessions, while the second consisted of data collected during the next three sessions. This setup investigated the efficacy of our model with data from the same platform.

\textit{Cross platform} For creating a cross-platform environment, we combined the from all six sessions for each platform. The training data came from platform $P_1$ while testing data from platform $P_2$, where $P_1$ and $P_2$ did not overlap. This generated $\binom{3}{2}$, i.e., six pairs of platforms. This setup investigated the efficacy of the proposed framework regardless of the platform. 

\textit{Combined-cross platform} We combined the corresponding dictionaries from the intended platforms to obtain the cross-platform training and testing data. For example, for training data, we combined the dictionaries from platform $P_1$ and $P_2$ while testing with the dictionaries created from data collected on $P_3$. This generated a total of three combinations. This setup investigated if our proposed framework performed better in cases with more training data. This setup showed the most consistent performance across the classifiers and experimental scenarios (see Section \ref{ResultsAndDiscussions}). 

\subsubsection{Verifiers}
We have employed modified versions of two established keystroke verification algorithms, viz. Similarity \cite{phoha2013method}, and Absolute Verifiers \cite{Gunetti2005} besides one state-of-the-art algorithm, the Instance-based Tail Area Density (ITAD) \cite{FastFreeText}. Although there exist numerous methods for modeling and classifying keystroke data \cite{shadman2023keystroke, stragapede2022typeformer, senerath2023behaveformer}, simple statistical methods have outperformed the state-of-the-art deep learning-based verifiers \cite{StatsVsDeepLearningKeystroke, shadman2023keystroke}. We describe these methods and present the variation of the associated algorithms we used in our study. 

\textbf{The Similarity (S) verifier:}
The Similarity Verifier employs Algorithm \ref{algo:Similarity} to compute the similarity score between an enrollment profile and a verification attempt \cite{Rahman2013}. We decided to use the median rather than the mean when computing features latency to reduce the impact of outliers. Unlike existing literature, which frames the Similarity Verifier algorithm in terms of dissimilarity, we express it in terms of similarity for clearer understanding. Thus, our formulation does not subtract the ratio of key matches to total features from one: $1-m/n$, where $m$ is the number of total feature matches, and $n$ is the number of total features compared. 
 
\begin{algorithm}[htp]
\caption{Calculate weighted similarity score}
\label{algo:Similarity}
\begin{algorithmic}
\REQUIRE Patterns \( A \), \( B \), with a set of common features \( C \).
\ENSURE Similarity score \( s \).
\IF{\( | C | == 0 \)}
\STATE \( a \gets 0 \) 
\RETURN a
\ENDIF
\STATE \( k \gets 0 \) \COMMENT{\( k \) for  feature matches}
\STATE \( t \gets 0 \) \COMMENT{\( t \) for total feature }
\FOR{each feature \( f \in C \)}
\STATE \( \widetilde{x} \gets \text{median}(A[f]) \)
\STATE \( \sigma \gets \begin{cases}
\sigma(A[f]) & \text{if } \sigma(A[f]) \text{ can be computed} \\
\frac{A[f]}{4} & \text{otherwise}
\end{cases} \)
\STATE \( v \gets 0 \) \COMMENT{\( v \) for feature value matches}
\STATE \( u \gets 0 \) \COMMENT{\( u \) for total feature values}
\FOR{each \( e \in B[f] \)}
\IF{\( \widetilde{x} - \sigma < e < \widetilde{x} + \sigma \)}
\STATE \( v \gets v + 1 \)
\ENDIF
\STATE \( u \gets u + 1 \)
\ENDFOR
\IF{\( \frac{v}{u} \leq 0.5 \)}
\STATE \( k \gets k + 1 \)
\ENDIF
\STATE \( t \gets t + 1 \)
\ENDFOR
\STATE \( s \gets \frac{k}{t} \)
\RETURN \( s \)
\end{algorithmic}
\end{algorithm}


\begin{figure*}[htp]
    \centering
    \includegraphics[width=6.9 in, height=1.85in]{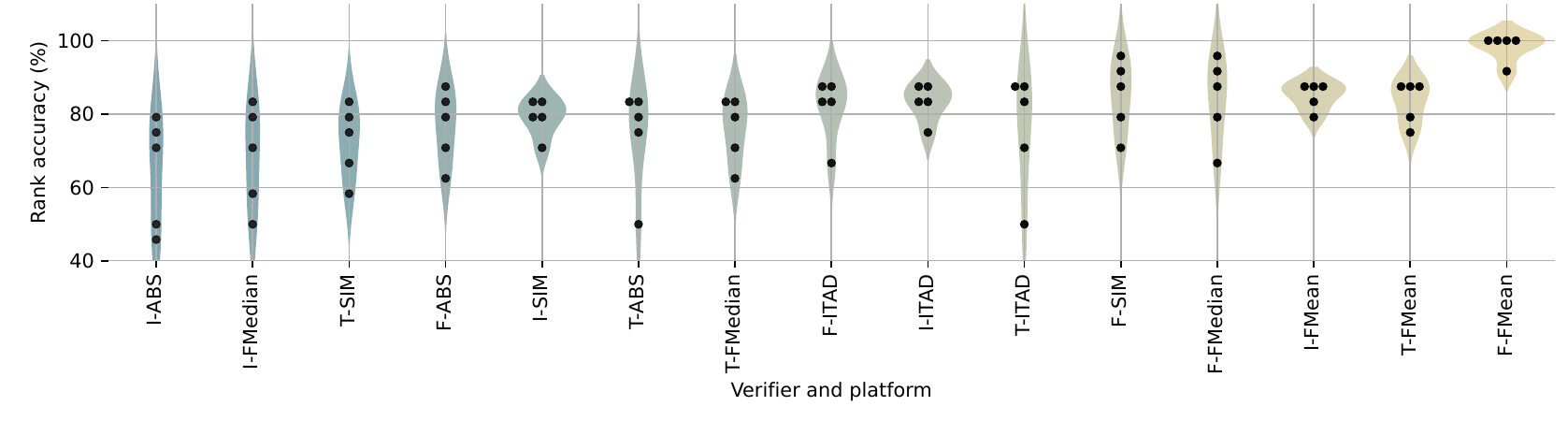}
    \caption{\textit{\textbf{Different sessions but same platform results:}} The violins represent the rank accuracies for different combinations of platform and verifier. F: Facebook, I: Instagram, T: Twitter, ABS: Absolute, SIM: Similarity, FMean: score level fusion using mean, and FMedian: score level fusion using the median. The naming convention, e.g., I-ABS, means the Performance of the Absolute verifier on Instagram data. Similarly, F-FMean indicates the Performance of the mean score-fusion-based verifier on Facebook data. }
    \label{fig:same-platform}
\end{figure*}
\begin{figure*}[htp]
    \centering
    \includegraphics[width=6.9 in, height=2in]{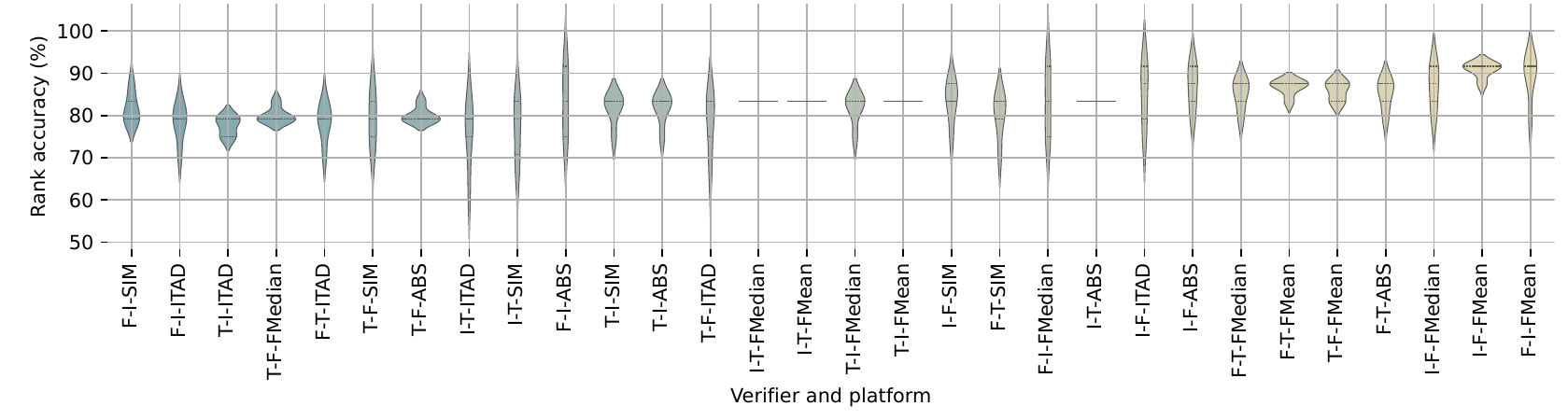}
    \caption{ \textit{\textbf{Cross-platform results:}} F: Facebook, I: Instagram, T: Twitter, ABS: Absolute, SIM: Similarity, FMean: score level fusion using mean, and FMedian: score level fusion using the median. The naming convention, e.g., F-T-ABS abbreviation, means that the model was trained on F, i.e., Facebook data, and tested on T, i.e., Twitter data using the Absolute verifier. }
    \label{fig:cross-platform}
\end{figure*}
\begin{figure*}[htp]
    \centering
    \includegraphics[width=6.9 in, height=2in]{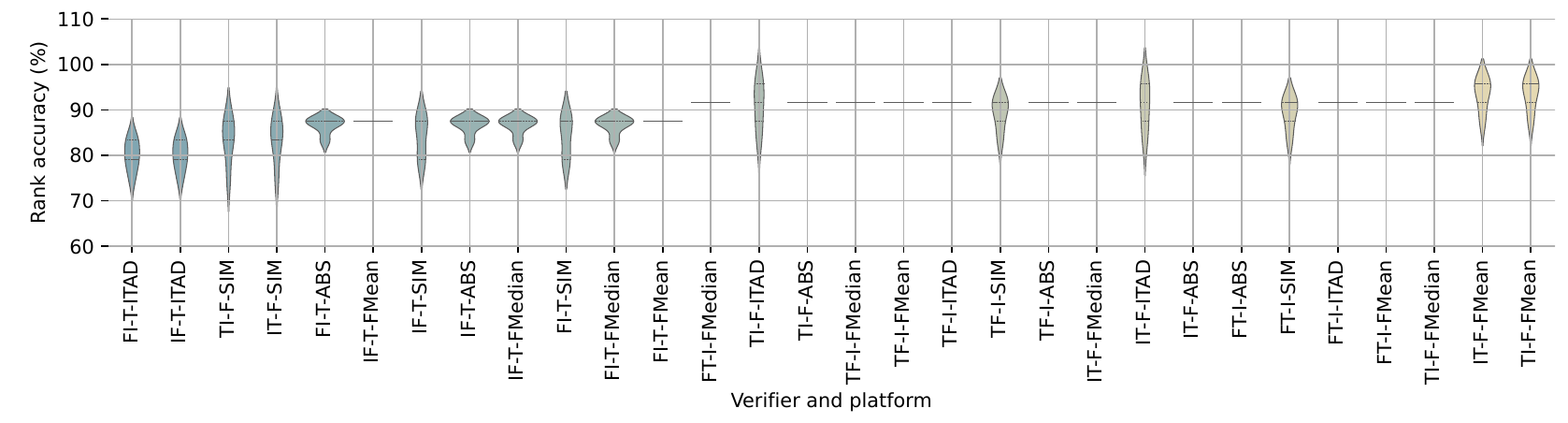}
    \caption{\textit{\textbf{Combined cross-platform:}} F: Facebook, I: Instagram, T: Twitter, ABS: Absolute, SIM: Similarity, FMean: score level fusion using mean, and FMedian: score level fusion using the median. The naming convention, e.g., FI-T-SIM abbreviation, means that the model was trained on combined data of Facebook and Instagram and tested on Twitter data using the Similarity verifier. }
    \label{fig:combined-cross}
\end{figure*}

\textbf{The Absolute (A) verifier:}
The Absolute Verifier employs Algorithm \ref{algo:Absolute} to compute the absolute match score between an enrollment profile and a verification attempt \cite{Gunetti2005}. Like the Similarity Verifier, we use the median instead of the mean to reduce the impact of latency outliers. For the same reason that the Similarity Verifier did not subtract the ratio of key matches to total features from one, we follow this approach in the Absolute Verifier. Additionally, we set the threshold to $1.5$ through trial and error.

\begin{algorithm}[!hbt]
\caption{Calculate absolute match score}
\label{algo:Absolute}
\begin{algorithmic}
\REQUIRE Patterns \( A \), \( B \), with a set of common features \( C \).
\ENSURE Absolute match score \( a \)

\IF{\( |C| == 0 \)}
    \STATE \( a \gets 0 \) \\
    \RETURN \( a \)
\ENDIF

\STATE \( T \gets 1.5 \) 

\STATE \( m \gets 0 \)  

\FOR{each \( f \) in \( C \)}
    \STATE \( \widetilde{x}_A \leftarrow \text{median}(A, f) \) \COMMENT{\( \widetilde{x}_A \): median of \( A \) for feature \( f \)}
    
    \STATE \( \widetilde{x}_B \leftarrow \text{median}(B, f) \) \COMMENT{\( \widetilde{x}_B \): median of \( B \) for feature \( f \)}
    
    \STATE \( r \leftarrow \text{max}(\widetilde{x}_A,
 \widetilde{x}_B) / \text{min}(\widetilde{x}_A, 
 \widetilde{x}_B) \)
    \IF{\( r \leq T \)}
        \STATE \( m \leftarrow m + 1 \)
    \ENDIF
\ENDFOR

\STATE \( a \gets \frac{m}{|C|} \)
\RETURN \( a \)
\end{algorithmic}
\end{algorithm}

\textbf{Instance-Based Tail Area Density (ITAD):}
The ITAD metric utilizes each common feature's tail area under the Probabilistic Density Function (PDF). The computation of the ITAD metric is described in Algorithm \ref{algo:CombinedITADv3}.

\begin{algorithm}[htp]
\caption{Calculate ITAD similarity score using ECDF}
\label{algo:CombinedITADv3}
\begin{algorithmic}
\REQUIRE Two patterns \( A \) and \( B \) with common features \( C \).
\ENSURE Similarity score \( i \).
\IF{\( | C | == 0 \)}
\RETURN 0
\ENDIF

\STATE \( Q\) $\gets$ \{\}

\FOR{each feature \( f \in C \)}
\STATE X $\gets A[f], Y \gets B[f]$  
\STATE \( \widetilde{x} \gets \text{median}(X) \)
\STATE $F(X) \gets$ ECDF(A[f])
\FOR{each value \( y \in Y \)}
\STATE \( p = F(X=y) \) \text{probability of X assuming value y}
\[
s = \begin{cases}
p & \text{if } y \leq \widetilde{x} \\
1 - p & \text{if } y > \widetilde{x}
\end{cases}
\]
\STATE append \( s \) to \( Q \).
\ENDFOR
\ENDFOR

$i \leftarrow \bar{Q}$

\RETURN $i$

\end{algorithmic}
\end{algorithm}

\subsubsection{Fusion}
We also implemented a score-level fusion function to improve our results further. Each verifier generated a matrix of size $n \times n$ as every user sample was compared with all other users, including its enrollment samples. To compute the resultant fused score matrix, we did the following: 

Let $A_{ij}, S_{ij}$, and $T_{ij}$ are the score matrices obtained for Absolute, Similarity, and ITAD verifiers, the $R_{ij}$ would be computed as follows: 
\begin{equation}
\begin{aligned}
\label{eq:ScoreFusion}
R_{ij} &= f(A_{ij}, S_{ij}, T_{ij}), \quad f \in \{\text{mean}, \text{median}\}, \\
& \quad i, j = 0, \ldots, n-1 
\end{aligned}
\end{equation}
Besides mean and median, we also investigated min and max-based score fusion, but they exhibited poorer performance than mean and median-based fusions. 

\subsection{Performance evaluation}
\label{PerformanceEval}
We compute the $k$-rank accuracy from the score matrices and report the same for $k$ ranging from $1$ to $5$. The $k$-rank accuracy means the prediction is correct if the correct class is in the top $k$ matches. 

\section{Results and discussion}
\label{ResultsAndDiscussions}
\subsection{Results}
Results for different experimental scenarios are presented in Figure \ref{fig:same-platform}, \ref{fig:cross-platform}, and \ref{fig:combined-cross}. Each violin represents the distribution of $k$-rank accuracies. The lowermost dot (minimum) in the violin plot indicates the rank-$1$ accuracy of the detector, while the upper-most dot (maximum) in the violin plot indicates the rank-5 accuracy of the detector. The violins are ordered according to the verifiers' median performance (accuracy) across datasets. It is worth noting that some violin plots are breaching the $100\%$ accuracy boundary because the kernel used for estimating the violin is a normal distribution, making the violin inflate beyond $100\%$ during estimation. The violins are ordered according to the verifiers' median performance (accuracy) across datasets for better readability.
As the results show, the score fusion verifiers perform the best, with the mean version performing better than the median. As a close second, the ITAD metric also performs well in most scenarios. The lack of plots in some scenarios, particularly in the combined cross-platform, speaks to the plateauing verifier performance after reaching a certain k-value.

\subsection{Discussion}
\label{Discussion}
Our results have shown that we have detected fake profiles on social network platforms with high accuracy using purely statistical and computationally efficient algorithms. 
We also investigated score-level fusion as our algorithms follow diverse decision-making paradigms. Mean and median performed better than individual algorithms of the four investigated fusion methods (mean, median, min, and max). We aim to investigate lexical and linguistic features such as Shannon Entropy and MATTR in the future \cite{cunningham2020, MLAndNLP}. Some evaluation scenarios showed no improvement after a certain value of $k$. A way to mitigate this effect is to explore more features. In addition, we would like to investigate the proposed approach on multiple devices and whether the learning transfers from one device \cite{MultiDeviceDetection} and one context \cite{Kumar2016BTAS, ContextAwarenessIJCB2017, ContextAwarenessCVPRW2014} to others.  

\section{Conclusion and future work}
\label{ConclusionAndFutureWork}
This paper has illustrated promise in keystroke-based fake profile detection in social networks. The experiment was conducted on a novel dataset explicitly designed for this task. Three widely studied and established keystroke verifiers were studied along with their score-level fusion. The fusion-based verifiers outperformed the individual verifiers. In future work, we aim to augment this dataset with additional users. We also aim to investigate the usefulness of additional such as linguistic features in enhancing the accuracy of the fake profile detection process. Plans for further research also include the exploration of multi-modal, multi-device, usage-context, and deep learning frameworks. 

{\small
\bibliography{references}
}
\end{document}